\documentclass[prd,aps,floats,a4paper,twocolumn,tightenlines,notitlepage]{revtex4}

\usepackage{epsfig}

\usepackage{amsmath}
\usepackage{amssymb}

\begin{document}
%

%
%

\def\lap{\nabla^2}

\def\gtt{G^{t}_{~t}}
\def\grr{G^{r}_{~r}}
\def\grz{G^{r}_{~z}}
\def\gzz{G^{z}_{~z}}
\def\gaa{G^{\theta}_{~\theta}}
\def\grrzz{(G^{r}_{~r}+G^{z}_{~z})}
\def\crz{G^{r}_{~z}}
\def\crrzz{(G^{r}_{~r}-G^{z}_{~z})}

\def\figuremode{\small}

%
%

\title{From Black Strings to Black Holes }

\author{Toby Wiseman} 

\email{T.A.J.Wiseman@damtp.cam.ac.uk} 

\affiliation{ DAMTP, CMS, University of Cambridge, Wilberforce Road, 
              Cambridge CB3 0WA, UK }
            
\date{November 2002}

%
\begin{abstract}
%
  
  Using recently developed numerical methods, we examine neutral
  compactified non-uniform black strings which connect to the
  Gregory-Laflamme critical point. By studying the geometry of the
  horizon we give evidence that this branch of solutions may connect
  to the black hole solutions, as conjectured by Kol. We find the
  geometry of the topology changing solution is likely to be nakedly
  singular at the point where the horizon radius is zero. We show that
  these solutions can all be expressed in the coordinate system
  discussed by Harmark and Obers.
 
  \vspace*{.3cm} {\small
\noindent DAMTP-2002-134 \hfill hep-th/0211028 }

%
\end{abstract}
%

\maketitle

%
\section{Introduction}
\label{sec:intro}
%

The existence of neutral non-uniform black string solutions was
originally postulated by Gregory and Laflamme when they discovered the
perturbative classical instability of the uniform strings
\cite{Gregory_Laflamme1,Gregory_Laflamme2,Gregory_Laflamme3,Gregory_Laflamme4}.
By compactifying the axis of rotational symmetry, the solutions can be
stabilised above a critical mass. At the critical mass, the marginal
mode is static indicating a new branch of solutions with spontaneously
broken translational invariance along the compact direction. The end
state of the decay for the unstable compactified uniform strings was
supposed to be a compact black hole solution, with spherical spatial
horizon topology. However, Horowitz and Maeda \cite{Horowitz_Maeda1,
  Horowitz} argued that the horizon can not `pinch off' in a finite
affine time, and concluded that the end state should instead be a
non-uniform string. The non-uniform solutions connected to the
Gregory-Laflamme (G-L) critical point were then constructed in third
order perturbation theory by Gubser \cite{Gubser}, who found that
their mass increased with non-uniformity, whilst the asymptotic
compactification radius is held constant. In a recent work
\cite{Wiseman3,Website} we have shown that the mass of these
non-uniform strings is in fact always greater than that of the
critical string.  This implies they can not be the product of decay of
an unstable uniform string, and the mysterious dynamics of the decay
remains an open question. Numerical work in progress may shed light on
the product, although the evolution time is not long enough at present
\cite{Horowitz, Choptuik}. Other interesting issues are links between
classical and thermodynamic stability for the uniform strings
\cite{Gubser_Mitra1,Gubser_Mitra2,Reall,Ross,Hubeny_Rangamani}, the
classical instability in related spacetimes
\cite{Gregory,Reall2,Kang,Gibbons_Hartnoll1}, and the phenomenological
implications these strings may have in theories with compact extra
dimensions \cite{Kol2,Kol3}.

Some analytic results are known for non-uniform strings
\cite{Horowitz_Maeda2,DeSmet}.  However, the Weyl ansatz which solves
4-dimensional static axisymmetric gravity does not generalise to
higher dimensions to give spherical symmetry about an axis \cite{Weyl,
  Myers, Emparan_Reall1, Emparan_Reall2}.  Interesting work by Harmark
and Obers \cite{Harmark_Obers} has conjectured an ansatz for the
general string/black-hole solution that reduces the number of unknown
metric functions from three to one, although they could not show its
consistency. Using this ansatz they conjectured a \emph{new} branch of
non-uniform string solutions, unrelated to those connected to the G-L
critical point.  We term these `conventional' non-uniform solutions
\emph{G-L strings}, and it is these that mostly concern us here. We
term the solutions conjectured to exist by Harmark and Obers the
\emph{H-O strings}.

Using Morse theory, Kol \cite{Kol1} has conjectured that the G-L
string branch of solutions should connect to the black hole branch
(assuming it actually exists - see \cite{Tanaka,Kaloper}) via a
spatial horizon topology changing solution.  In this Letter, we use
our numerical solutions for highly non-uniform G-L strings to study
the geometry of the horizon, as a function of the minimal sphere
radius, whilst keeping the asymptotic radius of compactification
fixed.  Following Gubser we will use the quantity, $\lambda =
\frac{1}{2} \left( \frac{\mathcal{R}_{max}}{\mathcal{R}_{min}} - 1
\right)$ where $\mathcal{R}_{max}$ is the maximum radius of a 3-sphere
on the horizon and $\mathcal{R}_{min}$ is the minimum, being the
radius at the `waist'. Thus $\lambda$ is zero for the uniform black
string.  We briefly review the numerical method, and then provide
evidence that as $\lambda \rightarrow \infty$ these solutions are
consistent with Kol's conjecture, that $\lambda = \infty$ is the
horizon topology changing solution, which connects with the black hole
branch.  Finally we demonstrate that our solutions can always be
written in the form of the Harmark and Obers ansatz, thereby showing
consistency of this ansatz, at least for G-L strings. We comment on
the implication of this for the existance of H-O strings.

%
\section{Constructing Solutions}
\label{sec:blackstring}
%

The problem of constructing compactified non-uniform string solutions
is an elliptic one. We wish to have an asymptotic product geometry of
flat space and an $S^1$. Near the axis of rotational symmetry we wish
to impose horizon boundary conditions, with a fixed amount of
`wiggliness'. In addition we wish the solutions to be periodic in the
axis direction.  The scale invariance of the vacuum Einstein equations
means that finding the solutions for a fixed $S^1$ size allows all
other solutions to be generated simply by scaling.

We first developed numerical methods in \cite{Wiseman1} to fully solve
the geometry of a star near its upper mass limit on a Randall-Sundrum
brane \cite{Randall_Sundrum1, Randall_Sundrum2}. Later in
\cite{Wiseman3} we extended these methods to the case of the black
string.  For technical reasons, we consider the 6 dimensional black
string solution, rather than the 5 dimensional one examined by Gubser.
As shown in \cite{Wiseman3}, these have the same qualitative
thermodynamic behaviour as 5 dimensional strings, to leading order in
$\lambda$. We take the metric,
\begin{equation}
ds^2 = - \frac{r^2}{m + r^2} e^{2 A} dt^2 + e^{2 B} ( dr^2 + dz^2 ) + e^{2 C}
( m + r^2 ) d\Omega^2_{3}
\label{eq:bs_metric}
\end{equation}
where $A, B, C$ are functions of $r, z$ and $z = [0,L]$ is an interval
as the string is wrapping an $S^1$. This form can always be
\emph{locally} chosen. When $A, B, C = 0$ this reduces to a uniform
black string.  As found in \cite{Wiseman1, Wiseman3}, the diagonal
form, and residual invariance under $r, z$ conformal transformations
results in several important simplifications. Firstly, the `interior'
Einstein equations, $\gtt$, $\gaa$ and $\grrzz$ take the form,
\begin{equation}
\lap X_i = \mathrm{src}_{X_i}
\label{eq:interior}
\end{equation}
for $X_i = \{ A, B, C \}$, where $\lap = \partial_r^2 + \partial_z^2$,
and the sources $\mathrm{src}_{X_i}$ depend non-linearly on all the
$X_j$, $\partial_r X_j$ and $\partial_z X_j$. The form of these
equations, or equivalently the Lagrangian from which they are derived,
allows standard elliptic numerical methods to be employed to solve for
$A, B, C$ given elliptic boundary data. Note that the superspace
metric has Lorentzian signature, so this is an extremisation problem,
rather than a minimisation one. The second crucial observation is that
the remaining Einstein equations, $\crz$ and $\crrzz$, weighted by the
measure $g = \sqrt{- \det{g_{\mu\nu}}}$, obey an elegant Cauchy-Riemann
relation,
\begin{eqnarray}
\partial_r \left( g \, \crz \right) +
    \partial_z \left( \frac{g}{2} \, \crrzz \right) & = 0 \nonumber \\
\partial_z \left( g \, \crz \right) -
    \partial_r \left( \frac{g}{2} \, \crrzz \right) & = 0
\label{eq:bianchi}
\end{eqnarray}
provided that the interior equations are satisfied. If, for example,
we satisfy $g \crz$ on all the boundaries, we need only satisfy $g
\crrzz$ at one point and both constraints must then be zero everywhere
in the interior. Asymptotically we choose $A, B, C \rightarrow 0$, and
thus $L$ is the radius of the compactification.  We also enforce the
$\crrzz$ constraint there, and the $\crz$ constraint is automatically
true as $z$ dependence decays exponentially at large $r$ with these
boundary conditions.  The third important feature resulting from the
$r, z$ conformal invariance is that we are free to choose the
boundaries, where we fix data, to be at any coordinate location. Thus
we need not include additional information to encode boundary
positions.  We take periodic boundary conditions at $z = 0, L$, and
choosing the horizon to be at $r = 0$, compatible with our choice of
background string solution, implies that $A, B, C$ will be finite at
the horizon.  Horizon boundary conditions are imposed by regularity.
We find \cite{Wiseman3} that we must impose $A_{,r} = 0, C_{,r} = 0$
\footnote{The $\crrzz$ constraint is then satisfied, and consequently
  $B_{,r} = 0$, as a result of the constraint structure
  \eqref{eq:bianchi}.}  and the constant temperature condition,
\begin{equation}
( \partial_z A - \partial_z B ) \mid_{r=0} = 0
\label{eq:horizon_const}
\end{equation}
which ensures $\crz$ is satisfied, and must be integrated along the
horizon, giving rise to one parameter, $B_{max}$ at $z = L$, as the
integration constant, that determines the deformation from the uniform
string. From our numerical solutions, we find a one-to-one relation
between $\lambda$ and $B_{max}$.

\begin{figure}[phtb]
  \psfig{file=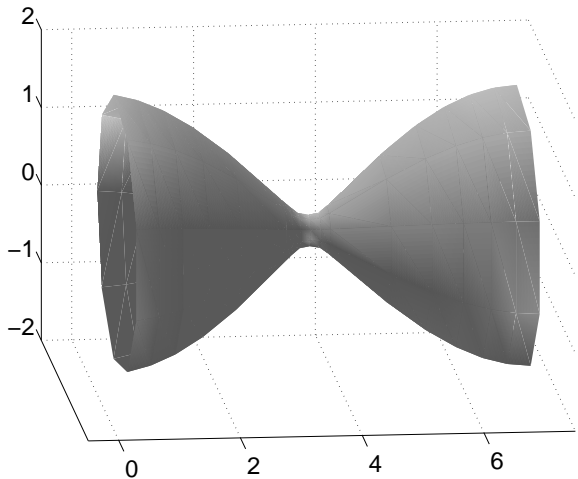,width=3in}
\caption{ 
  Horizon geometry of non-uniform string found with $\lambda \simeq
  4$. For the same asymptotic $S^1$ radius, the critical uniform
  string has unit horizon radius. This indicates that the large value
  of $\lambda$ is mainly due to a cycle shrinking, suggesting the
  possibility of a topology change for $\lambda \rightarrow \infty$.
\label{fig:horizon} 
}
\end{figure}

\begin{figure}[phtb]
  \psfig{file=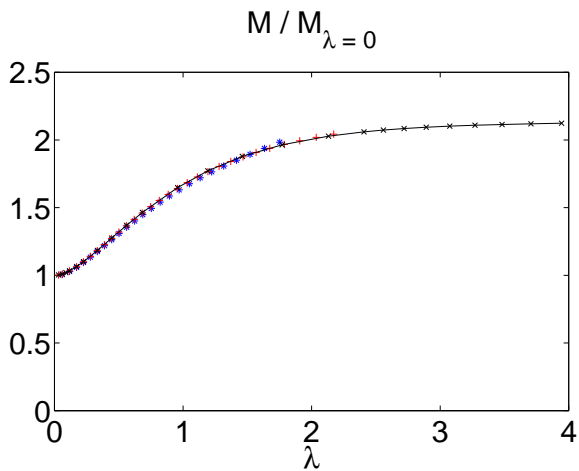,width=3in}
\caption{ 
  Plot of mass, normalised by the critical uniform string mass,
  against $\lambda$ for fixed asymptotic compactification radius. The
  data points indicate actual solutions. Three resolutions are shown
  superposed, the highest resolution allowing solutions to be found
  with $\lambda \simeq 4$.
\label{fig:mass} 
}
\end{figure}

Relaxation of the interior equations, with the boundary conditions
described above, is continued until a convergent solution is found.
Using elementary numerical methods and moderate resolution and
computer time, solutions were found for $\lambda \lesssim 4$. In
figure \ref{fig:horizon} we show the spatial horizon geometry of the
most non-uniform solution, embedded into $\mathbb{R}^3$.  Consistency
checks were performed in \cite{Wiseman3}; direct evaluation of the
constraints, comparison with third order perturbation theory at small
$\lambda$, varying $L$ (which gives scale equivalent solutions).  The
asymptotic ADM mass of the solutions is plotted in figure
\ref{fig:mass} against $\lambda$. This can be computed from the
asymptotic behaviour of the metric, or from integration of the First
Law, and the two give good agreement. For $\lambda \simeq 4$, the
difference in the values is approximately $5 \%$. For a detailed
discussion of errors, see \cite{Wiseman3}. It is difficult to assess
systematic errors in the method, but it is likely that other data
plotted here have similar magnitude errors. From this figure
\ref{fig:mass}, we concluded that the mass of these non-uniform G-L
strings is always larger than the critical string, and for $\lambda
\rightarrow \infty$ it tends to approximately twice the value of the
critical string.

%
\section{Kol's Conjecture}
\label{sec:kolconjecture}
%

\begin{figure}[htb]
\psfig{file=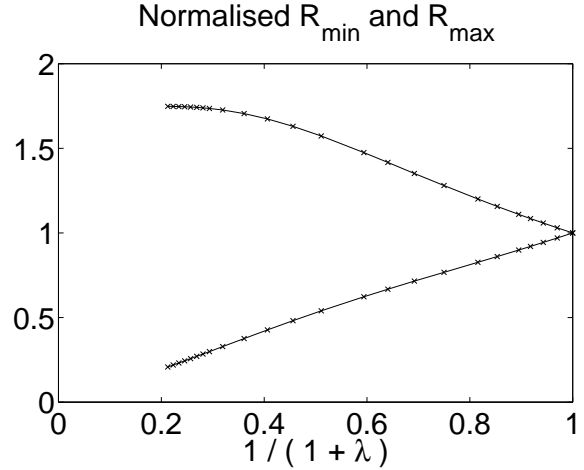,width=3in}
\caption{ 
  Plot of maximum and minimum horizon radius, normalised by the
  critical uniform string radius, against $1 / (1 + \lambda)$, for
  fixed asymptotic $S^1$ size. We see $R_{min} \propto 1 / \lambda$
  for $\lambda \rightarrow \infty$ and $R_{max}$ remains finite.
\label{fig:geom1} 
}
\end{figure}

We now consider \emph{how} $\lambda$ becomes large.  For our
solutions, the maximum value for the metric function $C$ occurs at $z
= 0$, and the minimum is at $z = L$, where $B = B_{max}$.  This gives
rise to the horizon 3-sphere radii $\mathcal{R}_{max}$ and
$\mathcal{R}_{min}$, plotted in figure \ref{fig:geom1} as a function
of $1 / (1 + \lambda)$, for fixed asymptotic $S^1$ radius, and
normalised by the critical uniform string horizon radius.  Clearly for
$\lambda$ to become large, either $\mathcal{R}_{min} \rightarrow 0$,
or $\mathcal{R}_{max} \rightarrow \infty$, or both. It is immediately
clear from the figure that $\lambda$ becomes large due to
$\mathcal{R}_{min}$ shrinking, apparently as $1 / \lambda$ for large
$\lambda$.  The earlier horizon embedding, figure \ref{fig:horizon},
graphically illustrates this, the critical uniform string with the
same asymptotic $S^1$ size having unit horizon radius. Thus we find
good evidence that $\mathcal{R}_{min} \rightarrow 0$ as $\lambda
\rightarrow \infty$.

Taking the black hole branch of solutions, for the same asymptotic
compactification radius, and increasing their mass from zero, the
black holes will appear less and less like 6 dimensional Schwarzschild
solutions.  At some point, the branch of solutions might cease, there
could be a point where the horizon topology changes
\cite{Harmark_Obers,Kol1}, the solutions might decompactify along the
axis, as occurs for the extremal charged black hole solutions, or the
black hole could become highly extended only in the radial direction
\cite{Myers,Horowitz2}.  If, as Kol suggests, there is a point of
topology change and the solutions connect to the G-L non-uniform
branch at $\lambda = \infty$, so $\mathcal{R}_{min} = 0$, we would
expect a \emph{finite} $\mathcal{R}_{max}$. On the other hand, if this
is not the case, the geometry of the G-L strings may do something
quite different as $\lambda \rightarrow \infty$. An example would be
if Harmark and Obers are correct, and it is a \emph{different} branch
of non-uniform solutions - the H-O strings - that connect to the black
hole solutions, rather than the G-L solutions.

We see from figure \ref{fig:geom1} that $\mathcal{R}_{max}$ does
appear to tend to a constant value as $\lambda \rightarrow \infty$.
Thus by considering the extrema of the horizon, we find consistency
with the idea that the black hole solutions could connect to the
non-uniform branch. Let us now consider other geometric quantities, to
see if pathologies develop that could spoil this picture. 

\begin{figure}[htb]
\psfig{file=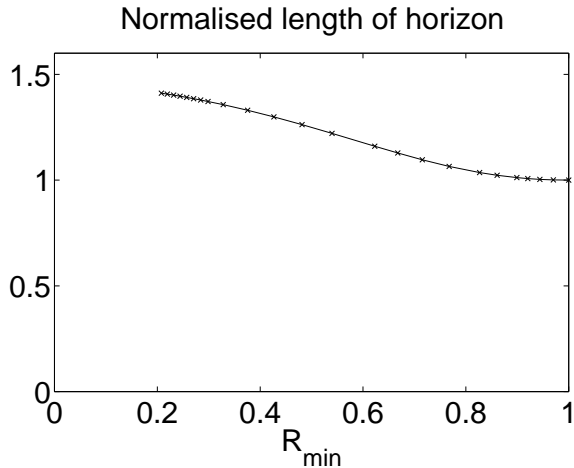,width=3in}
\caption{ 
  Plot of proper length of horizon against $R_{min}$, normalised by
  the critical uniform string value, for fixed asymptotic $S^1$
  radius.
\label{fig:geom2} 
}
\end{figure}

Firstly, whilst we fix the asymptotic length of the $S^1$, we do not a
priori know the proper length along the horizon, $L_{horiz}$, found by
integrating $e^{B}$ at $r = 0$.  We plot this in figure
\ref{fig:geom2}, and see that it increases, but appears consistent
with tending to a constant as $\mathcal{R}_{min} \rightarrow 0$, again
being compatible with the solution tending to the limiting black hole
solution which can only just `fit' into the circle. Note that another
possible option would have been finding this horizon length became
infinite in the limit, which obviously would not have been compatible
with Kol's conjecture.

\begin{figure}[htb]
\psfig{file=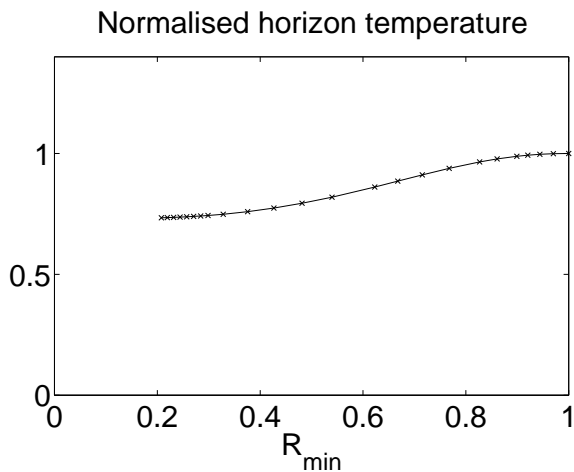,width=3in}
\caption{ 
  Plot of horizon temperature against $R_{min}$, normalised by same
  the value for the critical uniform string, for fixed asymptotic $S^1$ size.
\label{fig:geom3} 
}
\end{figure}

Thus we have considered coordinate invariant quantities associated
with the $B, C$ metric functions. The remaining metric function $A$
forms the coordinate invariant horizon temperature, as $e^{A-B}$.  As
shown in \cite{Wiseman3}, and in figure \ref{fig:geom3}, the
temperature again appears to tend to a constant.  This is
\emph{crucial} as we expect this temperature to remain finite if there
is a continuous interpolation between black hole and string solutions.

Note that finding finite values for both the above quantities is
encouraging from a numerical point of view, as at $z = L$, where the
horizon radius is a minimum, we find both $e^{A}$ and $e^{B}$ appear
to diverge as $\lambda \rightarrow \infty$.  Obviously we expect the
coordinate system to become singular at the topology changing point,
and it is good news that such coordinate invariant quantities remain
well behaved there.

To summarise, several geometric quantities appear to have limits as
$\mathcal{R}_{min} \rightarrow 0$ compatible with interpolating to
black hole solutions, as conjectured by Kol. We stress that this is by
no means a proof.  However, the conjecture could easily have been
falsified if any of the tested quantities had not behaved in a finite
way as $\lambda \rightarrow \infty$, so we do regard it as supporting
evidence. Assuming Kol's picure holds, the values of $R_{max}$,
$L_{horizon}$ and the horizon temperature in the above figures,
extrapolated to $\lambda \rightarrow \infty$, can then be taken as
testable \emph{predictions} for the same quantities measured in the
largest mass black hole solution. If, for example, our numerical
methods can be modified to explore the black hole branch, we may then
\emph{quantitatively} test whether the branches could join, even if
the topology changing solution can not be numerically constructed
explicitly.

%
\section{Horizon Geometry as $\lambda \rightarrow \infty$}
\label{sec:horizongeom}
%

We now consider the geometry at the minimum radius 3-sphere of the
horizon. A simple scalar measure of curvature is the Kretschmann
invariant, $K = R^{\alpha\beta\mu\nu} R_{\alpha\beta\mu\nu}$.  Since
we are interested in its behaviour at this extremal point of the
horizon, $r = 0$, $z = L$, and $\partial_{r,z} X_i = 0$ there, we may
expand to quadratic order in $r, \hat{z}$,
\begin{equation}
X_i \simeq x_{(0) i} + x_{(1) i} \hat{z}^2 + x_{(2) i} r^2
\label{eq:horiz_behaviour}
\end{equation}
where $x_{(j) i} = \{a_{(j)}, b_{(j)}, c_{(j)}\}$ for $X_i = \{A, B,
C\}$, and $\hat{z} = z+L$. Using the interior equations we may solve
$x_{(2) i}$ in terms of $x_{(0,1) i}$, and the constant temperature
condition, resulting from the constraints, yields $a_{(1)} = b_{(1)}$.
We find,
\begin{equation}
K = 48 \left( e^{ - 2 c_{(0)} } - \frac{9}{8} c_{(1)} e^{ - 2 b_{(0)} } \right)^2 + \frac{525}{4} \left( c_{(1)} e^{ - 2 b_{(0)} } \right)^2 
\label{eq:curvature}
\end{equation}
at the waist $r = 0$, $\hat{z} = 0$, choosing units where $m = 1$.  We
now know that for this minimal cycle, $c_{(0)} \rightarrow - \infty$
as $\lambda \rightarrow \infty$ (as $e^{c_{(0)}} = \mathcal{R}_{min}
\rightarrow 0$). From the form of the equation above, it is easy to
see that $K$ \emph{must} diverge in this limit, independent of the
behaviour of $b_{(0)}$ and $c_{(1)}$. Thus we have shown the curvature
at the waist must become singular, provided the solutions exist for
large $\lambda$ with the boundary conditions we impose, so that the
near waist behaviour \eqref{eq:horiz_behaviour} remains true. Assuming
that $A$ remains finite away from the horizon then implies the
$\lambda = \infty$ solution has a naked singularity at the waist. This
again agrees with Kol's picture, who has also suggested that this
nakedly singular geometry near the waist may be a cone metric, which
we aim to test in future work.

%
\section{The Harmark-Obers Ansatz}
\label{sec:harmark_obers}
%

Finally, we discuss the metric ansatz proposed by Harmark and Obers,
applied to 6-dimensional neutral black string/hole geometries, namely,
\begin{equation}
ds^2 = - \frac{\bar{r}^2}{r_0^2 + \bar{r}^2} dt^2 + e^{2 M} ( d\bar{r}^2 + e^{-6 K} d\bar{z}^2 ) + e^{2 K} ( r_0^2 + \bar{r}^2 ) d\Omega^2_{3}
\label{eq:HO_metric}
\end{equation}
where $M, K$ are functions of $\bar{r}, \bar{z}$, and $r_0$ is a
constant. In \cite{Harmark_Obers}, Harmark and Obers constructed this
ansatz based on a coordinate system that `interpolates' between the
black hole and uniform string solutions. They also find that $K$ can
be algebraically eliminated in terms of $M$. This therefore gives an
ansatz containing only one free function, $M$, rather than the usual 3
generally required by static axisymmetry, as in our metric
\eqref{eq:bs_metric}.

Of the five independent Einstein equations discussed above, one linear
combination vanishes, one is used to determine $K$, leaving 3
remaining equations for $M$. Harmark and Obers were unable to show
that these could be solved consistently, but gave suggestive evidence
by considering the consistency of the ansatz at the horizon, and to
second order in an asymptotic expansion in $\bar{r}$.  

In fact it is easy to show the 3 Einstein equations may be locally
consistent, due to the 2 independent components of the contracted
Bianchi identities. However, the question of global consistency
appears to depend on the explicit existence of solutions.  In this
section we exhibit the coordinate transformation that takes our metric
coordinate system \eqref{eq:bs_metric} into the Harmark-Obers
coordinate system.  Our metric contains sufficiently many metric
functions to parameterise the general solution locally, and thus this
coordinate transformation `locally' demonstrates the consistency of
the Harmark-Obers ansatz.  However, our numerical solutions indicate
that our coordinates are good globally for the G-L strings (with
finite $\lambda$), and therefore shows the `global' consistency of
their ansatz for these solutions.

Taking the coordinate transformation,
\begin{eqnarray}
\bar{r} & = f(r,z) \\ \nonumber   
\bar{z} & = h(r,z)
\end{eqnarray}
then gives the conditions,
\begin{eqnarray}
\label{eq:fh_eqns}
\partial_r f & = + e^{- 3 K} \partial_z h \\ \nonumber   
\partial_z f & = - e^{- 3 K} \partial_r h
\end{eqnarray}
for the transformation of \eqref{eq:bs_metric} to Harmark-Obers form.
In addition, we require that after the transformation the lapse is the
specific function in \eqref{eq:HO_metric}. The task is then to
understand what $f, h$ solve these equations, and consistently gives
only this $\bar{r}$ dependence in the lapse. For this lapse to take
the desired form, we would require,
\begin{equation}
f^2 = \frac{r_0^2 r^2 e^{2 A}}{m + r^2 (1 - e^{2 A})}
\end{equation}
To check whether this can be a solution to the coordinate
transformation conditions \eqref{eq:fh_eqns}, we may eliminate $h$ from
these to give,
\begin{equation}
\lap f + 3 ( \partial_r f \partial_r K + \partial_z f \partial_z K ) = 0
\end{equation}
Substituting our required $f$ above into this condition, gives a
second order equation for $A$ which is exactly the interior equation,
$\lap A = \mathrm{src}_{A}$ of \eqref{eq:interior}. Thus the $f$
required by the Harmark-Obers ansatz is indeed a consistent solution
of this coordinate transform, and thus we may transform our metric
into the Harmark-Obers form. 

We now consider the positions of the horizon, and periodic boundaries.
In our solutions, we have chosen them to be at $r = 0$ and $z = 0, L$,
respectively. We require $\partial_z X_i = 0$ at the periodic
boundaries. $A, B, C$ are finite on the horizon, and go to zero at
large $r$. Together, these facts imply that in the $\bar{r}, \bar{z}$
coordinates, the horizon is at $\bar{r} = 0$, and the periodic
boundaries are at constant $\bar{z}$. With a suitable choice of data
for $h$ in the transformation above we may pick $\bar{z} = 0$ to be
one periodic boundary.  To find the second periodic boundary location,
say $\bar{z} = \bar{L}$, one must then use $K$ and \eqref{eq:fh_eqns} to
give,
\begin{equation}
\partial_z h = \frac{1}{r_0^2} e^{A + 3 C} \left( m + r (m + r^2) \partial_r A \right)
\end{equation}
At large $r$, $A, C$ only have $r$ power law dependence, the $z$
dependence having decayed away exponentially. One integrates to find
$h(r,z) = \bar{z}$ at large $r$, which gives $\bar{L}$ at $z = L$. As
$A, B, C \rightarrow 0$ as $r \rightarrow \infty$ so $M, K$ also
vanish there and thus $\bar{L}$ is the proper asymptotic circle
length.  It is now important that we may choose $r_0$ so that $L =
\bar{L}$ ie. choosing $r_0$ correctly `undoes' the global scaling that
will generically occur in this coordinate transformation.  What is
crucial is that for fixed $m, L = \bar{L}$, this means that $r_0$ is a
function of the specific solution $A, B, C$, and therefore of our
parameter $\lambda$.

Plotting the function $f$ shows its $r$ derivative is everywhere
positive, within these boundaries, for the solutions available with
$\lambda < 4$. Then \eqref{eq:fh_eqns} similarly shows that the $z$
derivative of $h$ will be positive everywhere.  This implies that for
the G-L strings tested, no pathologies of the $\bar{r}, \bar{z}$ H-O
coordinate system arise in the interior.

Thus the Harmark-Obers ansatz appears to be entirely consistent for
the G-L branch of solutions. Indeed taking $r_0$ as the required
function of $\lambda$, the periodic boundaries are then at constant
$\bar{z} = 0, L$, and the horizon at $\bar{r} = 0$, which is the same
as considered in \cite{Harmark_Obers}. From inspection of the Einstein
equations resulting from the interpolating H-O ansatz, Harmark and
Obers proposed that the black hole solutions went through a topology
changing solution into the H-O strings, which they thought to be
\emph{distinct} from the G-L strings. They parameterised their
solutions exactly using $r_0$. Inverting the arguments above, we now
could think of our G-L strings as being parameterised by $r_0$ rather
than $\lambda$.

To summarise; we have shown that $\lambda \rightarrow \infty$ appears
to be a topology changing limit for the G-L strings, have shown they
can be consistently expressed in the H-O ansatz, with the same
boundary conditions and boundary locations as discussed by Harmark and
Obers, and furthermore have shown how their parameterisation of
solutions in terms of $r_0$ fits with the one in terms of $\lambda$.
In addition, Harmark and Obers also predict a nakedly singular
topology changing point, as Kol does, and as we give numerical
evidence for here. It is then intriguing to consider whether, rather
than being a new distinct branch of solutions, these H-O non-uniform
solutions are simply the G-L strings. This would certainly appear to
be the simpler outcome. Then Harmark and Obers argument for the
continuation of the black hole solutions into some non-uniform string
solution appears to be consistent with Kol's expectations.

If this is the case, the Harmark-Obers ansatz might provide important
clues in how to construct the G-L string solutions through the
topology changing point, where an interpolating coordinate system must
be devised which still allows elliptic numerical methods to be used.

%
\section{Conclusion}
\label{sec:conclusion}
%

Using our recently developed elliptic numerical methods, we present
evidence supporting Kol's conjecture, namely that the compactified G-L
nonuniform string solutions connect to the black hole branch of
solutions as $\lambda \rightarrow \infty$.  The evidence is based on
examining several geometric quantities which must remain finite as the
minimal horizon sphere shrinks to zero radius, if the conjecture is to
be true. If true, we may now make quantitative predictions about the
mass and geometry of the maximal black hole solution. We find the
curvature at the waist appears to diverge in the $\lambda \rightarrow
\infty$ limit, implying a naked singularity for the topology changing
solution.  We have also shown consistency of the Harmark-Obers ansatz
for these G-L string solutions, which we see are contained in this
ansatz. Following from this, we suggest that there are no new distinct
nonuniform H-O strings, as conjectured by Harmark and Obers, and the
nonuniform strings they consider are simply the G-L strings, tying in
their work with Kol's picture.

%
\section*{Acknowledgements}
%

I am grateful to Ruth Gregory, Gary Gibbons and Harvey Reall for
useful discussions. In particular, I thank Barak Kol and Troels
Harmark for discussions and also comments on this manuscript.  This
work was supported by Pembroke College, Cambridge, and computations
were performed on COSMOS at the National Cosmology Supercomputing
Centre in Cambridge.

%
%

%

\begin{thebibliography}{35}
\expandafter\ifx\csname natexlab\endcsname\relax\def\natexlab#1{#1}\fi
\expandafter\ifx\csname bibnamefont\endcsname\relax
  \def\bibnamefont#1{#1}\fi
\expandafter\ifx\csname bibfnamefont\endcsname\relax
  \def\bibfnamefont#1{#1}\fi
\expandafter\ifx\csname citenamefont\endcsname\relax
  \def\citenamefont#1{#1}\fi
\expandafter\ifx\csname url\endcsname\relax
  \def\url#1{\texttt{#1}}\fi
\expandafter\ifx\csname urlprefix\endcsname\relax\def\urlprefix{URL }\fi
\providecommand{\bibinfo}[2]{#2}
\providecommand{\eprint}[2][]{\url{#2}}

\bibitem[{\citenamefont{Gregory and Laflamme}(1993)}]{Gregory_Laflamme1}
\bibinfo{author}{\bibfnamefont{R.}~\bibnamefont{Gregory}} \bibnamefont{and}
  \bibinfo{author}{\bibfnamefont{R.}~\bibnamefont{Laflamme}},
  \bibinfo{journal}{Phys. Rev. Lett.} \textbf{\bibinfo{volume}{70}},
  \bibinfo{pages}{2837} (\bibinfo{year}{1993}),
  \eprint[http://arXiv.org/abs]{hep-th/9301052}.

\bibitem[{\citenamefont{Gregory and Laflamme}(1988)}]{Gregory_Laflamme2}
\bibinfo{author}{\bibfnamefont{R.}~\bibnamefont{Gregory}} \bibnamefont{and}
  \bibinfo{author}{\bibfnamefont{R.}~\bibnamefont{Laflamme}},
  \bibinfo{journal}{Phys. Rev.} \textbf{\bibinfo{volume}{D37}},
  \bibinfo{pages}{305} (\bibinfo{year}{1988}).

\bibitem[{\citenamefont{Gregory and Laflamme}(1994)}]{Gregory_Laflamme3}
\bibinfo{author}{\bibfnamefont{R.}~\bibnamefont{Gregory}} \bibnamefont{and}
  \bibinfo{author}{\bibfnamefont{R.}~\bibnamefont{Laflamme}},
  \bibinfo{journal}{Nucl. Phys.} \textbf{\bibinfo{volume}{B428}},
  \bibinfo{pages}{399} (\bibinfo{year}{1994}),
  \eprint[http://arXiv.org/abs]{hep-th/9404071}.

\bibitem[{\citenamefont{Gregory and Laflamme}(1995)}]{Gregory_Laflamme4}
\bibinfo{author}{\bibfnamefont{R.}~\bibnamefont{Gregory}} \bibnamefont{and}
  \bibinfo{author}{\bibfnamefont{R.}~\bibnamefont{Laflamme}},
  \bibinfo{journal}{Phys. Rev.} \textbf{\bibinfo{volume}{D51}},
  \bibinfo{pages}{305} (\bibinfo{year}{1995}),
  \eprint[http://arXiv.org/abs]{hep-th/9410050}.

\bibitem[{\citenamefont{Horowitz and Maeda}(2001)}]{Horowitz_Maeda1}
\bibinfo{author}{\bibfnamefont{G.}~\bibnamefont{Horowitz}} \bibnamefont{and}
  \bibinfo{author}{\bibfnamefont{K.}~\bibnamefont{Maeda}},
  \bibinfo{journal}{Phys. Rev. Lett.} \textbf{\bibinfo{volume}{87}},
  \bibinfo{pages}{131301} (\bibinfo{year}{2001}),
  \eprint[http://arXiv.org/abs]{hep-th/0105111}.

\bibitem[{\citenamefont{Horowitz}(2002)}]{Horowitz}
\bibinfo{author}{\bibfnamefont{G.}~\bibnamefont{Horowitz}},
  \emph{\bibinfo{title}{Playing with black strings}} (\bibinfo{year}{2002}),
  \eprint[http://arXiv.org/abs]{hep-th/0205069}.

\bibitem[{\citenamefont{Gubser}(2001)}]{Gubser}
\bibinfo{author}{\bibfnamefont{S.}~\bibnamefont{Gubser}},
  \emph{\bibinfo{title}{On non-uniform black branes}} (\bibinfo{year}{2001}),
  \eprint[http://arXiv.org/abs]{hep-th/0110193}.

\bibitem[{\citenamefont{Wiseman}()}]{Website}
\bibinfo{author}{\bibfnamefont{T.}~\bibnamefont{Wiseman}},
  \emph{\bibinfo{title}{http://www.damtp.cam.ac.uk/user/tajw2}}.

\bibitem[{\citenamefont{Wiseman}(2002{\natexlab{a}})}]{Wiseman3}
\bibinfo{author}{\bibfnamefont{T.}~\bibnamefont{Wiseman}},
  \emph{\bibinfo{title}{Static axisymmetric vacuum solutions and non-uniform
  black strings}} (\bibinfo{year}{2002}{\natexlab{a}}),
  \eprint[http://arXiv.org/abs]{hep-th/0209051}.

\bibitem[{\citenamefont{Choptuik et~al.}()\citenamefont{Choptuik, Lehner,
  Olabarrieta, Petryk, Pretorius, and Villegas}}]{Choptuik}
\bibinfo{author}{\bibfnamefont{M.}~\bibnamefont{Choptuik}},
  \bibinfo{author}{\bibfnamefont{L.}~\bibnamefont{Lehner}},
  \bibinfo{author}{\bibfnamefont{I.}~\bibnamefont{Olabarrieta}},
  \bibinfo{author}{\bibfnamefont{R.}~\bibnamefont{Petryk}},
  \bibinfo{author}{\bibfnamefont{F.}~\bibnamefont{Pretorius}},
  \bibnamefont{and} \bibinfo{author}{\bibfnamefont{H.}~\bibnamefont{Villegas}},
  \emph{\bibinfo{title}{To appear}}.

\bibitem[{\citenamefont{Gubser and Mitra}(2000)}]{Gubser_Mitra1}
\bibinfo{author}{\bibfnamefont{S.}~\bibnamefont{Gubser}} \bibnamefont{and}
  \bibinfo{author}{\bibfnamefont{I.}~\bibnamefont{Mitra}},
  \emph{\bibinfo{title}{Instability of charged black holes in anti-de sitter
  space}} (\bibinfo{year}{2000}),
  \eprint[http://arXiv.org/abs]{hep-th/0009126}.

\bibitem[{\citenamefont{Gubser and Mitra}(2001)}]{Gubser_Mitra2}
\bibinfo{author}{\bibfnamefont{S.}~\bibnamefont{Gubser}} \bibnamefont{and}
  \bibinfo{author}{\bibfnamefont{I.}~\bibnamefont{Mitra}},
  \bibinfo{journal}{JHEP} \textbf{\bibinfo{volume}{08}}, \bibinfo{pages}{018}
  (\bibinfo{year}{2001}), \eprint[http://arXiv.org/abs]{hep-th/0011127}.

\bibitem[{\citenamefont{Reall}(2001)}]{Reall}
\bibinfo{author}{\bibfnamefont{H.}~\bibnamefont{Reall}},
  \bibinfo{journal}{Phys. Rev.} \textbf{\bibinfo{volume}{D64}},
  \bibinfo{pages}{044005} (\bibinfo{year}{2001}),
  \eprint[http://arXiv.org/abs]{hep-th/0104071}.

\bibitem[{\citenamefont{Gregory and Ross}(2001)}]{Ross}
\bibinfo{author}{\bibfnamefont{J.}~\bibnamefont{Gregory}} \bibnamefont{and}
  \bibinfo{author}{\bibfnamefont{S.}~\bibnamefont{Ross}},
  \bibinfo{journal}{Phys. Rev.} \textbf{\bibinfo{volume}{D64}},
  \bibinfo{pages}{124006} (\bibinfo{year}{2001}),
  \eprint[http://arXiv.org/abs]{hep-th/0106220}.

\bibitem[{\citenamefont{Hubeny and Rangamani}(2002)}]{Hubeny_Rangamani}
\bibinfo{author}{\bibfnamefont{V.}~\bibnamefont{Hubeny}} \bibnamefont{and}
  \bibinfo{author}{\bibfnamefont{M.}~\bibnamefont{Rangamani}},
  \bibinfo{journal}{JHEP} \textbf{\bibinfo{volume}{05}}, \bibinfo{pages}{027}
  (\bibinfo{year}{2002}), \eprint[http://arXiv.org/abs]{hep-th/0202189}.

\bibitem[{\citenamefont{Gregory}(2000)}]{Gregory}
\bibinfo{author}{\bibfnamefont{R.}~\bibnamefont{Gregory}},
  \bibinfo{journal}{Class. Quant. Grav.} \textbf{\bibinfo{volume}{17}},
  \bibinfo{pages}{L125} (\bibinfo{year}{2000}),
  \eprint[http://arXiv.org/abs]{hep-th/0004101}.

\bibitem[{\citenamefont{Chamblin et~al.}(2000)\citenamefont{Chamblin, Hawking,
  and Reall}}]{Reall2}
\bibinfo{author}{\bibfnamefont{A.}~\bibnamefont{Chamblin}},
  \bibinfo{author}{\bibfnamefont{S.}~\bibnamefont{Hawking}}, \bibnamefont{and}
  \bibinfo{author}{\bibfnamefont{H.}~\bibnamefont{Reall}},
  \bibinfo{journal}{Phys. Rev.} \textbf{\bibinfo{volume}{D61}},
  \bibinfo{pages}{065007} (\bibinfo{year}{2000}),
  \eprint[http://arXiv.org/abs]{hep-th/9909205}.

\bibitem[{\citenamefont{Hirayama and Kang}(2001)}]{Kang}
\bibinfo{author}{\bibfnamefont{T.}~\bibnamefont{Hirayama}} \bibnamefont{and}
  \bibinfo{author}{\bibfnamefont{G.}~\bibnamefont{Kang}},
  \bibinfo{journal}{Phys. Rev.} \textbf{\bibinfo{volume}{D64}},
  \bibinfo{pages}{064010} (\bibinfo{year}{2001}),
  \eprint[http://arXiv.org/abs]{hep-th/0104213}.

\bibitem[{\citenamefont{Gibbons and Hartnoll}(2002)}]{Gibbons_Hartnoll1}
\bibinfo{author}{\bibfnamefont{G.}~\bibnamefont{Gibbons}} \bibnamefont{and}
  \bibinfo{author}{\bibfnamefont{S.}~\bibnamefont{Hartnoll}},
  \emph{\bibinfo{title}{A gravitational instability in higher dimensions}}
  (\bibinfo{year}{2002}), \eprint[http://arXiv.org/abs]{hep-th/0206202}.

\bibitem[{\citenamefont{Kol}(2002{\natexlab{a}})}]{Kol2}
\bibinfo{author}{\bibfnamefont{B.}~\bibnamefont{Kol}},
  \emph{\bibinfo{title}{Explosive black hole fission and fusion in large extra
  dimensions}} (\bibinfo{year}{2002}{\natexlab{a}}),
  \eprint[http://arXiv.org/abs]{hep-ph/0207037}.

\bibitem[{\citenamefont{Kol}(2002{\natexlab{b}})}]{Kol3}
\bibinfo{author}{\bibfnamefont{B.}~\bibnamefont{Kol}},
  \emph{\bibinfo{title}{Speculative generalization of black hole uniqueness to
  higher dimensions}} (\bibinfo{year}{2002}{\natexlab{b}}),
  \eprint[http://arXiv.org/abs]{hep-th/0208056}.

\bibitem[{\citenamefont{Horowitz and Maeda}(2002)}]{Horowitz_Maeda2}
\bibinfo{author}{\bibfnamefont{G.}~\bibnamefont{Horowitz}} \bibnamefont{and}
  \bibinfo{author}{\bibfnamefont{K.}~\bibnamefont{Maeda}},
  \bibinfo{journal}{Phys. Rev.} \textbf{\bibinfo{volume}{D65}},
  \bibinfo{pages}{104028} (\bibinfo{year}{2002}),
  \eprint[http://arXiv.org/abs]{hep-th/0201241}.

\bibitem[{\citenamefont{Smet}(2002)}]{DeSmet}
\bibinfo{author}{\bibfnamefont{P.~D.} \bibnamefont{Smet}},
  \emph{\bibinfo{title}{Black holes on cylinders are not algebraically
  special}} (\bibinfo{year}{2002}),
  \eprint[http://arXiv.org/abs]{hep-th/0206106}.

\bibitem[{\citenamefont{Weyl}(1917)}]{Weyl}
\bibinfo{author}{\bibfnamefont{H.}~\bibnamefont{Weyl}}, \bibinfo{journal}{Ann.
  Phys. (Leipzig)} \textbf{\bibinfo{volume}{54}}, \bibinfo{pages}{117}
  (\bibinfo{year}{1917}).

\bibitem[{\citenamefont{Emparan and
  Reall}(2002{\natexlab{a}})}]{Emparan_Reall1}
\bibinfo{author}{\bibfnamefont{R.}~\bibnamefont{Emparan}} \bibnamefont{and}
  \bibinfo{author}{\bibfnamefont{H.}~\bibnamefont{Reall}},
  \bibinfo{journal}{Phys. Rev.} \textbf{\bibinfo{volume}{D65}},
  \bibinfo{pages}{084025} (\bibinfo{year}{2002}{\natexlab{a}}),
  \eprint[http://arXiv.org/abs]{hep-th/0110258}.

\bibitem[{\citenamefont{Emparan and
  Reall}(2002{\natexlab{b}})}]{Emparan_Reall2}
\bibinfo{author}{\bibfnamefont{R.}~\bibnamefont{Emparan}} \bibnamefont{and}
  \bibinfo{author}{\bibfnamefont{H.}~\bibnamefont{Reall}},
  \bibinfo{journal}{Phys. Rev. Lett.} \textbf{\bibinfo{volume}{88}},
  \bibinfo{pages}{101101} (\bibinfo{year}{2002}{\natexlab{b}}),
  \eprint[http://arXiv.org/abs]{hep-th/0110260}.

\bibitem[{\citenamefont{Myers}(1987)}]{Myers}
\bibinfo{author}{\bibfnamefont{R.}~\bibnamefont{Myers}},
  \bibinfo{journal}{Phys. Rev.} \textbf{\bibinfo{volume}{D35}},
  \bibinfo{pages}{455} (\bibinfo{year}{1987}).

\bibitem[{\citenamefont{Harmark and Obers}(2002)}]{Harmark_Obers}
\bibinfo{author}{\bibfnamefont{T.}~\bibnamefont{Harmark}} \bibnamefont{and}
  \bibinfo{author}{\bibfnamefont{N.}~\bibnamefont{Obers}},
  \bibinfo{journal}{JHEP} \textbf{\bibinfo{volume}{05}}, \bibinfo{pages}{032}
  (\bibinfo{year}{2002}), \eprint[http://arXiv.org/abs]{hep-th/0204047}.

\bibitem[{\citenamefont{Kol}(2002{\natexlab{c}})}]{Kol1}
\bibinfo{author}{\bibfnamefont{B.}~\bibnamefont{Kol}},
  \emph{\bibinfo{title}{Topology change in general relativity and the
  black-hole black-string transition}} (\bibinfo{year}{2002}{\natexlab{c}}),
  \eprint[http://arXiv.org/abs]{hep-th/0206220}.

\bibitem[{\citenamefont{Tanaka}(2002)}]{Tanaka}
\bibinfo{author}{\bibfnamefont{T.}~\bibnamefont{Tanaka}},
  \emph{\bibinfo{title}{Classical black hole evaporation in randall-sundrum
  infinite braneworld}} (\bibinfo{year}{2002}),
  \eprint[http://arXiv.org/abs]{hep-th/0203082}.

\bibitem[{\citenamefont{Emparan et~al.}(2002)\citenamefont{Emparan, Fabbri, and
  Kaloper}}]{Kaloper}
\bibinfo{author}{\bibfnamefont{R.}~\bibnamefont{Emparan}},
  \bibinfo{author}{\bibfnamefont{A.}~\bibnamefont{Fabbri}}, \bibnamefont{and}
  \bibinfo{author}{\bibfnamefont{N.}~\bibnamefont{Kaloper}},
  \emph{\bibinfo{title}{Quantum black holes as holograms in ads braneworlds}}
  (\bibinfo{year}{2002}), \eprint[http://arXiv.org/abs]{hep-th/0206155}.

\bibitem[{\citenamefont{Wiseman}(2002{\natexlab{b}})}]{Wiseman1}
\bibinfo{author}{\bibfnamefont{T.}~\bibnamefont{Wiseman}},
  \bibinfo{journal}{Phys. Rev.} \textbf{\bibinfo{volume}{D65}},
  \bibinfo{pages}{124007} (\bibinfo{year}{2002}{\natexlab{b}}),
  \eprint[http://arXiv.org/abs]{hep-th/0111057}.

\bibitem[{\citenamefont{Randall and
  Sundrum}(1999{\natexlab{a}})}]{Randall_Sundrum1}
\bibinfo{author}{\bibfnamefont{L.}~\bibnamefont{Randall}} \bibnamefont{and}
  \bibinfo{author}{\bibfnamefont{R.}~\bibnamefont{Sundrum}},
  \bibinfo{journal}{Phys. Rev. Lett.} \textbf{\bibinfo{volume}{83}},
  \bibinfo{pages}{3370} (\bibinfo{year}{1999}{\natexlab{a}}),
  \eprint[http://arXiv.org/abs]{hep-ph/9905221}.

\bibitem[{\citenamefont{Randall and
  Sundrum}(1999{\natexlab{b}})}]{Randall_Sundrum2}
\bibinfo{author}{\bibfnamefont{L.}~\bibnamefont{Randall}} \bibnamefont{and}
  \bibinfo{author}{\bibfnamefont{R.}~\bibnamefont{Sundrum}},
  \bibinfo{journal}{Phys. Rev. Lett.} \textbf{\bibinfo{volume}{83}},
  \bibinfo{pages}{4690} (\bibinfo{year}{1999}{\natexlab{b}}),
  \eprint[http://arXiv.org/abs]{hep-th/9906064}.

\bibitem[{\citenamefont{Elvang and Horowitz}(2002)}]{Horowitz2}
\bibinfo{author}{\bibfnamefont{H.}~\bibnamefont{Elvang}} \bibnamefont{and}
  \bibinfo{author}{\bibfnamefont{G.}~\bibnamefont{Horowitz}},
  \emph{\bibinfo{title}{When black holes meet kaluza-klein bubbles}}
  (\bibinfo{year}{2002}), \eprint[http://arXiv.org/abs]{hep-th/0210303}.

\end{thebibliography}
\end{document}